\documentclass{article}
\usepackage{ijcai22}

\usepackage{times}
\usepackage{color}
\usepackage{soul}
\usepackage{url}
\usepackage[hidelinks]{hyperref}
\usepackage[utf8]{inputenc}
\usepackage[small]{caption}
\usepackage{graphicx}
\usepackage{amsmath}
\usepackage{amsthm}
\usepackage{booktabs}
\usepackage{algorithm}
\usepackage{algorithmic}
\usepackage{multirow}
\usepackage{bbding}

\title{From Quantum Graph Computing to Quantum Graph Learning: A Survey}

\author{
Yehui Tang$^{1}$\and
Junchi Yan$^1$\footnote{
Correspondence author is Junchi Yan.}\and
Edwin Hancock$^2$
\affiliations
$^1$Department of CSE, and MoE Key Lab of Artificial Intelligence, 
Shanghai Jiao Tong University\\
$^2$Department of Computer Science, University of York\\
\emails  
\{yehuitang,\ yanjunchi\}@sjtu.edu.cn \quad edwin.hancock@york.ac.uk}

\begin{document}

\maketitle

\begin{abstract}
Quantum computing (QC) is a new computational paradigm whose foundations relate to quantum physics. Notable progress has been made, driving the birth of a series of quantum-based algorithms that take advantage of quantum computational power. In this paper, we provide a targeted survey of the development of QC for graph-related tasks. We first elaborate the correlations between quantum mechanics and graph theory to show that quantum computers are able to generate useful solutions that can not be produced by classical systems efficiently for some problems related to graphs. For its practicability and wide-applicability, we give a brief review of typical graph learning techniques designed for various tasks. Inspired by these powerful methods, we note that advanced quantum algorithms have been proposed for characterizing the graph structures. We give a snapshot of quantum graph learning where expectations serve as a catalyst for subsequent research. We further discuss the challenges of using quantum algorithms in graph learning, and future directions towards more flexible and versatile quantum graph learning solvers.
\end{abstract}
\vspace{-3pt}
\section{Introduction}
\vspace{-3pt}   
Quantum computing (QC), whose foundations relate to quantum physics, harnesses the computational power derived from the properties of quantum mechanics to perform calculations. Quantum computing has attracted wide attention from both industry and academia for its promising prospective of exponentially accelerating the traditional approaches \cite{liu2021rigorous,zlokapa2021quantum} as well as naturally exploring the physical systems \cite{preskill2018quantum,hegade2021shortcuts}.

Benefiting from the unique characteristics of quantum mechanics, which are totally different from the classical systems, such as superposition, interference, and entanglement, quantum computing has potential superiority in many calculation tasks. The most remarkable step is taken by  \cite{shor1994algorithms} that demonstrates two mathematical problems - integer factorization and discrete logarithm - can be solved efficiently by using quantum computers, whereas no efficient classical method is known. Furthermore,  \cite{grover1996fast} points to another valuable evidence, showing the quantum advantage of searching and retrieving target values in unstructured databases quadratic faster than the most effective classical method.

The potential advantages of quantum computers arouse researchers' curiosity in solving the problem of graph theory. Early works concentrate on exploiting the quantum computational power to produce effective solutions to graph theoretical problems. We classify such quantum algorithms as quantum graph computing. Two families of graph theoretical problems have been studied extensively. One family relates to graph representations of combinatorial optimization problems which are thought to be exponentially hard for classical computers \cite{moylett2017quantum}. Another family comprises extracting patterns and finding hidden structures within a number of graphs \cite{chang2018quantum}. Most of these problems are NP-hard if not even harder, or practically intractable for large-scale settings for the exact solution. A number of quantum based algorithms focus on reformulating some of these problems into computationally solvable forms of quantum computers \cite{calude2017qubo}. Besides, these algorithms try to use the associated properties of quantum mechanics to solve problems efficiently and find the exact solution.

More recently modern graph learning techniques has revolutionized graph analytics tasks such as node classification \cite{kipf2017semi} and graph classification \cite{chen2020convolutional}. Meanwhile, fueled by the advent of quantum hardware and the high-performance quantum simulation frameworks, an increasing amount of literature on quantum machine learning has been proposed in recent years, showing the quantum potential of both acceleration of the computational process and improvement of the performance. With the power of data, quantum machine learning has the potential of finding atypical but useful patterns that classical systems are not considered to be able to generate effectively \cite{huang2021power}. It seems an attractive direction towards combining the expressive ability of classical graph learning techniques with the power of quantum computing. We call it quantum graph learning. In this paper, we give a snapshot of quantum graph learning and hope that it can inspire more in-depth research.
%Some advanced contributions have taken the first step. 

%However, their review lacks a theoretical discussion for near-term quantum computing.
There are reviews on quantum computing and quantum machine learning. \cite{montanaro2016quantum,biamonte2017quantum} overview quantum algorithms that have the potential speed-ups. They mainly review idealized quantum algorithms that efficiently perform linear algebraic operations. \cite{adedoyin2018quantum} conducts a survey focused on the gate-model quantum computer. \cite{preskill2018quantum} reports quantum algorithms that can be executed on the current quantum hardware and the future potential of quantum computing is discussed. \cite{dunjko2018machine} investigates the interaction between machine learning, artificial intelligence and quantum mechanics. \cite{cerezo2020variational} limits the scope within the quantum algorithms based on parameterized quantum circuits. In summary, existing surveys have not fully reflected the state-of-the-art development of quantum computing, especially for graph algorithms. Our survey focuses on quantum algorithms on graphs, from quantum graph computing to quantum graph learning, which we believe is an emerging field.
%, for researchers who want to enter this emerging field
\vspace{-3pt}
\section{Quantum Graph Computing}
\vspace{-3pt}
There are three common paths intersecting quantum computing and graph theory. First, some works consider an underlying graph structure consisting of nodes connected by edges and perform quantum evolution on this structure, in which an important quantum paradigm is Hamiltonian encoding. Second, the relationship between the transition rules of a graph and the randomness of the quantum representation is explored from the point of view of utilizing the superposition and entanglement of quantum states to characterize the properties of graphs, e.g., quantum random walks. In this case, the quantum states preserve the features of nodes and edges of the original graph as well as its topology information. Another way to build quantum graph solvers is the employment of quantum search, which, in most cases, provides quadratic acceleration over traditional exhaustive search methods. Here we discuss the synergy of these three areas.
%from the bidirectional perspective
\vspace{-3pt}
\subsection{Hamiltonian Encoding based Solvers}
\vspace{-3pt}
One approach to developing universal quantum computing is adiabatic quantum computing (AQC), which relies on the adiabatic theorem that evaluates the continuous-time evolution of a quantum system. To describe the dynamics of an arbitrary quantum system, a class of Hamiltonians is defined
\begin{equation}
\label{eq:ising}
    H = \sum_{i<j}J_{ij}\sigma_{i}^{z}\sigma_{j}^{z} + \sum_{i}h_i\sigma_{i}^{z} + \sum_{i}g_i\sigma_{i}^{x},
\end{equation}
where $\sigma^{z}$, $\sigma^{x}$ are Pauli matrices. Broadly speaking, Eq.~\ref{eq:ising} can be viewed as a transverse-field Ising model and such model is QMA-complete \cite{albash2018adiabatic}. The first two terms push the quantum bit (qubit) to be either $|0\rangle$ or $|1\rangle$, whereas the last pushes the state to a superposition. An outside measurement leads to the collapse from superposition to a deterministic state. These foundations give the ability for quantum computers to handle certain problems that can not be efficiently solved classically. To implement such models in a real physical process, quantum annealing (QA) captures the relaxation of the adiabatic conditions and controls the continuous-time evolution at finite temperature and in open environments. Machines performing QA at the hardware level attempt to minimize the energy determined by the Hamiltonian, and the output of an anneal is a low-energy ground state, which consists of an Ising spin for each qubit where the eigenvalues $\{+1,-1\}$ of Pauli matrix $\sigma^{z}$ correspond to the binary constraint of the properties of classical Ising model \cite{ushijima2017graph}. This is closely related to the quadratic unconstrained binary optimization (QUBO) problem - a combinatorial optimization problem that can be applied to many graph theoretical problems. 

Many NP-hard problems can be expressed as the minimization of the energy of the Hamiltonian as the form of Eq.~\ref{eq:ising} e.g. graph partitioning, graph coloring, vertex cover and clique finding \cite{lucas2014ising}. Therefore, varieties of contributions concentrate on relaxing and reformulating these graph-related combinatorial optimization problems into the QUBO formulation and utilizing QA to quickly approach the exact solution. The method in \cite{chams1987some} converts the graph coloring problem to graph partitioning and employs simulated annealing to screen the better solution of the problem. Different annealing schemes are exploited in \cite{silva2020mapping}. These schemes dominate traditional techniques for certain types of graphs with the cost of long computing time.
%Quantum annealing, whose dynamics is determined by the strength of the transverse field, exhibits the quantum advantage mainly since the amplitudes of all states are changed in parallel \cite{santoro2006optimization}.
To transform a set of constraints to the energy minimization problem, penalty terms associated with the constraints should be added to the QUBO quadratic expression \cite{silva2020mapping}. Besides, reduction and approximation are necessary for space efficient. Their work demonstrates that the quantum annealer can more reliably approach the optimal solution and produce better heuristics for graph coloring under specific settings.

Existing quantum hardware can only control a relatively small number of qubits, which is far from the size for large-scale data processing. In addition, it is difficult to fully embed real-world optimization problems on quantum devices as a consequence of poor connectivity between arbitrary couples of qubits. As a result, the qubit connectivity in quantum hardware rarely matches the QUBO form described by the underlying graph structure. To overcome these problems, the authors of \cite{bass2018heterogeneous} present a heterogeneous algorithm that uses classical co-processing to preprocess the primitive problem and randomly selects a large number of possible solutions, after which the reduced form of the max clique problem corresponding to the largest possible solutions can be encoded into the quantum annealer for accelerating the searching process. 
In \cite{pelofske2019solving}, a decomposition alternative divides a big input graph into multiple smaller subgraphs that fit the quantum annealer to further improve the adaptability.
Other applications also show the benefits of using Hamiltonian to encode and solve graph related problems including graph isomorphism problem \cite{calude2017qubo,ushijima2017graph} and vertex cover problem \cite{pelofske2019solving}. Likewise, recent works demonstrate quantum annealing has potential advantages over classical approaches ranging from optimization \cite{brady2021optimal} to machine learning \cite{nath2021review}. However, it is difficult to ensure the final quantum state to be the ground state, meaning that the final solution is probably not optimal. Besides, with the increase of the size of the system and the complexity of the problem, it seems reasonable to expect the annealing time to scale exponentially for problems with exponentially small gaps, especially for NP-hard problems \cite{albash2021comparing}. 
\vspace{-3pt}
\subsection{Quantum Random Walks based Solvers}
\vspace{-3pt}
Another way to represent the topological information of graphs in quantum information is quantum random walks (QRWs). Motivated by the widespread use of classical random walks, QRWs view a graph as a collection of nodes in either real or complex space, among which the correlations between two nodes are indicated by edges. The Walker evolves in a quantum mechanical manner over time by characterizing its initial distribution in terms of the amplitudes of quantum states. Compared with classical random walks, where the stochastic transition matrix determines the random patterns, the randomness of QRWs depends on the reversible unitary translation. No internal information could be obtained until the final states are measured. Readers are referred to  \cite{kempe2003quantum,venegas2012quantum} for introductions on QRWs. Here, we provide a brief overview of the two types of QRWs with typical progress on solving graph related tasks.

\paragraph{Discrete Time Quantum Random Walks} One way to intersperse quantum effects with random walks is the discrete time quantum random walks (DT-QRWs). Let $\mathcal{H}_{P}$ be the Hilbert space spanned by the positions of the nodes, and $\mathcal{H}_{C}$ be the `coin’-space whose dimension is usually equivalent to the maximal degree of the graph \cite{kempe2003quantum}. The state of the whole graph can be described by the space $\mathcal{H}=\mathcal{H}_{C} \otimes \mathcal{H}_{P}$. Suppose that a unitary operator $C$ determines the possibility of the diffusing direction, and $S$ indicates the conditional transition of the system, the DT-QRWs of step $T$ is defined as the transformation $U^T$, and $U$ is written as $U = S \cdot (C \otimes I)$, where $I$ is the identity matrix \cite{venegas2012quantum}. By repeating the succession of unitary translation $U$, i.e., transforming the state $|\psi_{t}\rangle = U|\psi_{t-1}\rangle$ iteratively with timestamp $t$, the distribution of the walker can be encoded in the final state. Measuring the coin-register of the walk in the computational basis states will output the classical information of the probability distribution.

\paragraph{Continuous Time Quantum Random Walks} Another way to model the quantum diffusing is the continuous time quantum random walks (CT-QRWs). For the general classical random walk in the form of the differential equation, the probability distribution at time $t$ of node $i$ is:
$
\frac{d p_{i}(t)}{d t}=-\sum_{j} H_{i, j} p_{j}(t),
$
where matrix $H$ is an analogue to the classical transition matrix, whereas each entry of $H$ is the possibility of jumping from node $i$ to node $j$. Solving the equation we obtain $\vec{p}(t)=e^{-Ht}\vec{p}(0)$, and \cite{farhi1998quantum} extends this concept to quantum case by using the Hamiltonian to establish the continuous evolution $U(t)=e^{-iHt}$. If we start in some initial state $|\psi_{0}\rangle$, evolve it under $U$ for a time $t$ and measure the positions of the resulting state we obtain a probability distribution over the nodes of the graph \cite{kempe2003quantum}. In contract to the classical random walks, quantum diffusion is a reversible process and it does not converge to a stationary distribution in general, and an average distribution called {\em Cesaro limit} $\vec{c}^t$ is introduced to obtain a stationary distribution: $\vec{c}_{i}^{t}=\frac{1}{t} \sum_{s=1}^{t} \vec{p}_{i}^{s}$.

\paragraph{Overview of QRWs-based Solvers}
These ideas of quantum diffusion are generalized to develop varieties of graph solvers for graph theoretical problems. QRWs often induce an asymmetric probability distribution mainly due to their intrinsic interference pattern and unusual collapse characteristics. It is evident that the interference pattern of QRWs is much more intricate than the Gaussian obtained in the classical case \cite{kempe2003quantum}. The first attempt to develop an exponential separation of classical and quantum random walks is given by \cite{childs2002example}. They construct a binary glue tree and demonstrate that DT-QRWs are able to penetrate the graph in polynomial time. \cite{childs2003quantum} develops an algorithm based on DT-QRWs for more general subgraph finding problems, i.e., L-subset distinctness in polynomial time for a small size of the subgraph. 
%Furthermore, they generalize their ideas and make a statement that certain graph traversal problems can be solved exponentially faster by using the quantum computer to perform the CT-QRWs. 
More intricate tasks including graph matching and graph distinction can also be handled using the quantum inference of QRWs \cite{emms2009graphd,emms2009graphc}.
Although there are many correlations between CT-QRWs and DT-QRWs, the authors of \cite{kempe2005discrete} find that in the hypercube case, the hitting time of DT-QRWs is exponentially faster than that of classical random walks, whereas the CT-QRWs do not converge to the uniform distribution at all. The relationship between the transition pattern of CT-QRWs and the spectrum of a regular graph is revealed by \cite{ren2011quantum}, which characterizes the capacity of the Ihara zeta function in distinguishing graphs. It is shown in \cite{emms2009graphe} that node embeddings produced by the hitting time associated with the CT-QRWs tend to capture more structural information.
%\cite{magniez2007quantum} proposes a method for tackling the triangle finding problem by only using the logarithmic number of qubits. 

\subsection{Quantum Search based Solvers}
% One of the important applications of QRWs is quantum search. There are similarities between Grover’s search algorithm and the quantum random walk search \cite{shenvi2003quantum}. An elegant way to implement quantum search is proposed by \cite{grover1997quantum}. 
The quantum search algorithm, also known as Grover's algorithm, is first proposed by \cite{grover1996fast} at the theoretical level. This algorithm incorporates a superposition of all possible states related to the problem. Then a unitary operator is used to change the amplitudes of each state while maximizing the amplitudes of the desired state which can be extracted by measuring. It is especially suitable for the application scenarios where a set of specific solutions are searched and retrieved in a huge number of candidates. For the function whose output size is $N$, the quantum search algorithm finds the desired input value by using just $O(\sqrt{N})$ evaluations of the function with high probability. In particular, many graph theoretical problems are NP-hard if not even tougher, and the search space grows exponentially with the problem's size. It is reasonable to handle these problems in terms of quantum computing power.

It is shown in \cite{durr2006quantum} that quantum search algorithm decreases the query complexity further for certain graph problems including spanning tree, connectivity, strong connectivity and single source shortest path. In \cite{hillery2010searching}, the search is performed by quantum random walks to find the marked clique of a complete graph. A novel quantum search architecture \cite{vcerny1993quantum} is developed to solve the travelling salesman problem (TSP). However, \cite{greenwood2001finding} points out this scheme does not mean quantum computers can solve arbitrary NP-hard problems in polynomial time since the number of qubits to represent the superposition states is astronomical. An efficient quantum approach proposed by \cite{srinivasan2018efficient} combines the quantum phase estimation algorithm \cite{kitaev1996quantum} with the quantum search algorithm to solve the TSP. It provides a quadratic speedup over the classical brute force method. A breakthrough to achieve quantum speedups for TSP and several NP-hard problems presented by \cite{ambainis2019quantum} integrates Grover's algorithm with classical dynamic programming. Despite the fact that quantum search has the potential to solve various complex graph related problems, developing a quantum algorithm that promises to be faster than the best classical algorithm remains difficult, as classical techniques do not always rely on exhaustive search.
%\cite{miyamoto2020quantum} applies this approach to the minimum Steiner tree problem. 
\vspace{-3pt}
\section{Classic Graph Learning}
\vspace{-3pt}
%Though the graph analysis over graph theory has been widely studied for a long time, it is still changeable to develop an end-to-end learning model for practical applications. 
%Graphs are credited with the ability to represent more general unstructured information in the real world, such as biology, social networks and linguistic \cite{goyal2018graph}. 
% \cite{kriege2020survey,wu2020comprehensive}
We provide background on classic graph learning. It in general attempts to assign a vector representation to each of the (sub)graphs, preserving both structural information and node features. 
%These steps can be either combined to perform an end-to-end learning process, or they can be enhanced further by training a neural network to a specific task.% We briefly review the classic graph learning methods in this section.
%and introduce quantum-based graph learning methods in Section \ref{sec:qgl}.
\vspace{-3pt}
\subsection{Factorization based Embedding Approaches}
\vspace{-3pt}
By representing the graph's edges and connected nodes as low-dimensional vectors that preserve global properties of the graph, the subsequent graph analytics tasks can be easily addressed by employing mature machine learning algorithms \cite{goyal2018graph}. Here we introduce the factorization based approaches and discuss their pros and cons.
% such as classification, clustering, and visualization

\textbf{Matrix Factorization based Approaches} There are a variety of graph learning approaches that represent the correlations between nodes by factorizing the matrix that contains the graph information to obtain the embedding. Different tasks require the factorization of matrices with different properties. Graph Laplacian eigenmaps \cite{belkin2001laplacian} constructs a low-dimensional representation for each node while keeping the smoothness of the distinctness of connected nodes. 
% Concretely, for the graph with $n$ nodes, the eigenvectors corresponding to the $d$ smallest eigenvalues of the normalized Laplacian matrix corresponds to the $d$-dimensional representation vectors of all nodes. 
To circumvent the loss of local topology information, a Cauchy graph embedding method developed by \cite{luo2011cauchy} is employed to preserve the similarity relationships of the original graph data, and introduce a more effective objective function to emphasize the similarity efficacy. However, the factorization of the matrix of the graph with massive nodes often requires huge computing resources. \cite{ahmed2013distributed} proposes a factorization technique that relies on graph partitioning to enhance scalability.
%of the factorization at the expense of some latent error. 

\textbf{Random Walk based Approaches} Random walk statistics is widely used to capture the graph properties. The latent representation of the node that captures the local structure information is condensed into the low-dimensional feature vector by performing a diffusing process. DeepWalk \cite{perozzi2014deepwalk} generates the training data in terms of compressing the graph structure into a text like corpus using random walk, and trains the graph learner to maximize the occurrence probability of neighbors in the walk. Node2vec \cite{grover2016node2vec} employs biased random walk to make a trade-off between breadth-first (BFS) and depth-first (DFS) graph searches, resulting in higher-quality and more informative embeddings than DeepWalk. \cite{qiu2018network} shows that many random walk based approaches also perform implicit matrix factorization, since the requirement of the specific adjacent matrix or Laplacian matrix before performing learning. Most of these methods are inherently transductive.
\vspace{-3pt}
\subsection{Graph Kernel Methods}
\vspace{-3pt}
Inspired by the kernel methods which utilize a linear classifier to solve non-linear problems, graph kernel methods have been widely used for graph-level tasks, e.g., classification and clustering. They directly compare the structures of the subgraphs by defining a user-specific similarity metric or a meaningful distance measurement on graphs. In contrast to the factorization based approaches representing the graph information as low-dimensional vectors, graph kernel methods characterize graph features implicitly in a high dimensional space. It involves performing pairwise comparisons between local substructures centered at every node. 
% For simple example, given two graphs $G$ with the set of nodes $\mathcal{V}$ and $G'$ with $\mathcal{V}'$, the graph kernel
% \begin{equation}
% K\left(G, G^{\prime}\right)=\sum_{u \in \mathcal{V}} \sum_{u^{\prime} \in \mathcal{V}^{\prime}} k\left(l_{G}(u), l_{G^{\prime}}\left(u^{\prime}\right)\right)
% \end{equation}
% It sums similar local patterns centered at the node $u$ and node $u'$ denoted by $l_{G}(u)$ and $l_{G'}(u')$ evaluated by kernel $k(\cdot,\cdot)$. 

Most early graph kernel works belong to the family of R-convolution kernels \cite{haussler1999convolution}. 
% The variants involve the random walk kernels that concentrate mainly on counting matching walks in the two input graphs \cite{zhang2018retgk}, the subtree kernels that counts the number of common subtree patterns in two graphs \cite{shervashidze2009fast}, and the path kernels that transform graphs into shortest paths to estimate their similarity \cite{borgwardt2005shortest}. 
% \cite{nikolentzos2021graph}\cite{xu2018powerful}
More recently, \cite{tian2019rethinking} shows a kernel-based framework to produce hidden representations of nodes, bridging the gap between graph kernel methods and graph neural networks. \cite{chen2020convolutional} combines the message passing mechanism with the graph kernel and constructs a convolutional kernel network to represent the graph effectively. 
The high computational complexity of graph kernel methods suppresses their applicability. While it is still valuable to study the theoretical framework behind the graph learning approaches using graph kernel methods due to their inherent transparency and interpretability. 
%They suggest building a finite-dimensional embedding for approximating the implicitly infinite representation of the graph to alleviate the scalability issues. 

\vspace{-3pt}
\subsection{Graph Neural Networks}
\vspace{-3pt}
%Inspired by the success of convolutional neural networks (CNNs), recurrent neural networks (RNNs), new generalizations and definitions of important operations have been developed to manage the complexity of graph data. In particular, 
For graph learning, Graph neural networks (GNNs) are a type of deep learning method for extracting the pattern of graph structural data, which dominate the literature.
%, leading to a transformative change to graph learning tasks.

\textbf{Recurrent Graph Neural Networks} The initial trial of the GNNs is recurrent graph neural networks whose foundations relate to the information diffusion mechanism. In the prototype~\cite{scarselli2008graph} of recurrent graph neural networks, the node hidden embedding is updated by the adjacent neighbors.
%\begin{equation}
%h_{v}^{(k)}=\sum_{u \in \mathcal{N}(v)} f\left(x_{v}, x_{u}, h_{u}^{(k-1)}\right)
%\end{equation}
%where $x_{v}$ is the original attribute of node $v$, $h_{v}^{(k)}$ denotes the hidden feature of node $v$ at the $k$-th iteration (except $h_{v}^{(0)}$ denotes the original attribute such that $h_{v}^{(0)}=x_{v}$), and $\mathcal{N}(v)$ is the adjacent neighbors of node $v$ and $f(\cdot)$ is a parameterized function. 
This process continues until a stable equilibrium is reached. A gated recurrent unit (GRU) is used as a recurrent function in a gated graph neural network (GGNN) \cite{li2015gated}, restricting the recurrence to a few steps.

%\cite{li2015gated} improves the efficiency by removing the convergence constraint to fix the number of recurrence steps. \cite{dai2018learning} handles the computational issue by the recurrent mechanism via introducing stochasticity.

\textbf{Convolutional Graph Neural Networks} A number of convolutional graph neural networks can be categorized into a general framework named Message Passing Neural Network (MPNN) \cite{gilmer2017neural}. The forward pass of these models has two phases - an aggregation phase and an update phase - that run iteratively to let information propagate further. The aggregation phase serves to detect (multiple) patterns in multiple sub-regions of the input graph, while the update phase generally consists of pooing functions that collapse the hidden features of interrelated sub-regions into a new vector representation. Some studies perform graph convolution operation building upon the signal processing on graphs \cite{shuman2013emerging}. 

\begin{table*}[tb!]
\centering
\resizebox{.9\linewidth}{!}{
\begin{tabular}{|c|c|c|c|c|c|c|c|}
\hline
Category                            & Method & Attribute & Embedding & Input & Layer  & Readout    & Application          \\ \hline
\multirow{4}{*}{QK-based}        & QJSK  \cite{bai2015quantum} & \XSolidBrush & \Checkmark     & C     & Q \& C & Tomography & Graph Classification \\ \cline{2-8} 
                                    & GBSK~\cite{schuld2020measuring}   & \XSolidBrush & \Checkmark     & C     & Q \& C & Estimation & Graph Classification \\ \cline{2-8} 
                                    & SFGK~\cite{kishi2021graph}   & \XSolidBrush & \Checkmark     & C     & Q \& C & Swap Test  & Graph Classification \\ \cline{2-8} 
                                    & QEK~\cite{henry2021quantum}    & \XSolidBrush & \Checkmark     & C     & Q \& C & Tomography & Graph Classification \\ \hline
\multirow{5}{*}{Shallow Circuit} & QGNN~\cite{verdon2019quantum}   & \XSolidBrush & \Checkmark     & Q     & Q      & Tomography & Graph Isomorphism    \\ \cline{2-8} 
                                    & QGCN~\cite{zheng2021quantum}   & \Checkmark   & \XSolidBrush   & Syn.  & Q \& C & Estimation & Graph Classification \\ \cline{2-8} 
                                    & DQGNN~\cite{ai2022decompositional}  & \Checkmark   & \XSolidBrush   & C     & Q \& C & Tomography & Graph Classification \\ \cline{2-8} 
                                    & EQGC~\cite{mernyei2021equivariant}   & \XSolidBrush & \Checkmark     & Syn.  & Q \& C & Estimation & Graph Classification \\ \cline{2-8} 
                                    & QNN~\cite{beer2021quantum}    & \XSolidBrush & \Checkmark     & Q     & Q      & Estimation & Network Embedding    \\ \hline
\multirow{5}{*}{Hybrid Deep}     & QWNN~\cite{dernbach2018quantum}   & \Checkmark   & \Checkmark     & C     & Q \& C & Tomography & Node Classification  \\ \cline{2-8} 
                                    & QSGCNN~\cite{bai2021learning} & \Checkmark   & \Checkmark     & C     & Q \& C & Tomography & Graph Classification \\ \cline{2-8} 
                                    & QSCNN~\cite{zhang2019quantum}  & \Checkmark   & \Checkmark     & C     & Q \& C & Tomography & Node Classification  \\ \cline{2-8} 
                                    & QGCNN~\cite{chen2021hybrid}  & \Checkmark   & \XSolidBrush   & C     & Q \& C & Estimation & Graph Classification \\ \cline{2-8} 
                                    & HQGNN~\cite{tuysuz2021hybrid}  & \XSolidBrush & \Checkmark     & C     & Q \& C & Estimation & Link Prediction      \\ \hline
\end{tabular}
}
\vspace{-5pt}
\caption{Comparison between quantum graph learning methods and their application. Although most of these methods employ the topology embedding that incorporates structural information into the quantum representation, node attributes are not considered in some research. The input data is classical (C), quantum (Q) or synthetic (Syn.). For the methods with classical or synthetic inputs, the classical layer is inevitably introduced to preprocess the graph data or assist the quantum computer to update the model parameters. The readout operation is the interaction transforming the quantum information into the classical expression, where the tomography may require an exponentially large number of measurements, whereas a small amount of measurements is necessary for estimation of the probability outcomes and swap test.}
\label{tab:qgl}
\vspace{-5pt}
\end{table*}

\vspace{-3pt}
\section{Quantum Graph Learning}
\label{sec:qgl}
\vspace{-3pt}
Recently many quantum machine learning algorithms have been proposed that either promise quantum speed-ups over their classical counterparts \cite{liu2021rigorous,zlokapa2021quantum}, or have the potential of finding atypical but useful patterns that classical systems are not considered to be able to generate effectively \cite{cong2019quantum,huang2021power}. With the development of quantum devices and high-performance quantum simulation frameworks, there is an increasing interest in building novel quantum algorithms to help solve near-term applications for practical problems. A variety of quantum analogues to the classical machine learning models have been proposed. Advanced contributions have taken the first step. For instance, \cite{harrow2009quantum} proposes an exponentially fast algorithm for efficiently solving linear equations. Based on this, the quantum support vector machine (QSVM) presented by \cite{rebentrost2014quantum} can handle binary classification problems on a quantum computer with complexity logarithmic in the size of the vectors and the number of training examples. The practical implementation of QSVM for recognizing handwritten characters is completed by \cite{li2015experimental}. To realize more intricate tasks, the quantum convolutional neural network (QCNN) \cite{cong2019quantum} generates excitement around the possibility of efficiently analyzing quantum data via performing convolutional and pooling operations in quantum systems.
%It is shown in \cite{oh2020tutorial} that QCNN works properly in image classification for the MNIST dataset. 

Although quantum algorithms have the potential to tackle graph problems efficiently, quantum computing for graph learning is still in its early stages. The literature is relatively sparse and lacks formal rationale for the model selections. In the following, we show some progress of leveraging quantum physics to extract graph structural information, bringing up new possibilities for quantum computing applications. The main characteristics and differences of these methods are summarized in Tab.~\ref{tab:qgl}.
%Here we give a snapshot of quantum graph learning where expectations serve as a catalyst for subsequent research.
\vspace{-3pt}
\subsection{Quantum Kernel based Graph Learning}
\vspace{-3pt}
Similar to the classical graph kernel methods, quantum kernels (QK) on graphs aim to decompose the graph into substructures and compare the similarity between each pair of substructures specified in terms of their quantum representations. The Quantum Jensen–Shannon Graph Kernel (QJSK) \cite{bai2015quantum} infers the graph properties using the density matrix description constructed by CT-QRWs. The authors develop an aligned version of the QJSK to preserve the permutation invariance and more correspondence information between pairs of nodes in the graph. Additionally, they generalize their ideas by probing the graph structure using the DT-QRWs \cite{bai2017quantum}. In \cite{schuld2020measuring}, the authors present the potential of encoding the graph information in the quantum Hilbert space using the derivation of Gaussian Boson Samplers Kernel (GBSK). The information of all possible subgraphs can be obtained by sampling instead of counting the appearance of specific substructures classically. A specific kernel encoding the feature of all subgraphs (SFGK) \cite{kishi2021graph} investigates how to utilize the power of quantum superposition to encode every subgraph into a feature space. The major disadvantage of their model is that the similarity between the superposition states is hard to estimate, thereby limiting the model capacity. The Hamiltonian encoding approach is further exploited in the Quantum Evolution Kernel (QEK) \cite{henry2021quantum} to represent the topology of the input graph. A particular graph kernel generated by performing the quantum evolution followed by a carefully chosen observable is able to characterize graphs in a real quantum computer.
\vspace{-3pt}
\subsection{Shallow Circuit Graph Learning}
\vspace{-3pt}
Most classic graph learning techniques are motivated by graph-free models such as kernel methods and CNNs. New generalizations and definitions of important operations have been developed to handle the complexity of graph data \cite{wu2020comprehensive}. As a result, some studies use quantum computing to efficiently duplicate procedures resulting from classic graph learning. It seems to be essential to build effective embedding approaches that encode the classical data into their quantum representations while keeping as much of the original information as possible \cite{huang2021power,schuld2021supervised}. Besides, the embedding phase contributes the most nonlinear transformations \cite{schuld2019quantum} and is the basis of the quantum speed-ups claimed by many quantum machine learning algorithms \cite{biamonte2017quantum}. Therefore, new methods should be developed to handle embedding efficacy. For learning on graphs, the permutation invariance of node orderings should be also carefully considered when transforming original graphs into their quantum expressions. The shallow circuit learning method is a kind of quantum algorithm based on the current noisy intermediate-scale quantum (NISQ) devices, which encodes the original data into their quantum representations and approximates the objective function by adjusting the learnable gate circuit parameters \cite{schuld2021supervised}. Generally the shallow circuit can be divided into two parts where the embedding circuit is responsible for encoding the data into quantum Hilbert space and the following parameterized processing circuit is used to extract hidden features from the feature space.

\cite{verdon2019quantum} introduces a quantum graph neural network (QGNN) that treats the interaction of qubits as the nodes connected by edges. The entire graph thereby can be expressed by a quadratic Hamiltonian, which implies that we can think about quantum circuits with graph-theoretic properties. Their numerical results limited to small-scale experiments show several potential applications such as spectral clustering and graph isomorphism classification. While more recently, it has been successfully applied to graphs with node features in \cite{zheng2021quantum,ai2022decompositional}. However, their methods suffer the computational cost as a result of quantum tomography, resulting in the massive resource overhead. \cite{mernyei2021equivariant} constructs an equivalent quantum graph circuit (EQGC) whose topology preserves permutation invariance of the input graph. But the number of the qubits scales linearly with the number of nodes, and the prediction accuracy closely depends on the fidelity of the measurements. An alternative approach devised by \cite{beer2021quantum} uses qubits as neurons to characterize the graph as quantum states. They design a quantum neural network (QNN) trained on a well-designed loss evaluated by the fidelity of quantum representations of connected nodes.
\vspace{-3pt}
\subsection{Hybrid Deep Graph Learning}
\vspace{-3pt}
While GNNs represent state-of-the-art classical machine learning for a range of benchmark tasks on graphs, there is no clear proof of quantum advantages for tasks on classical graph-related datasets in extant quantum neural networks. Therefore, some experts pin their hopes on developing approaches that combine the expressiveness of the classical learning methods with the power of quantum modules to improve the performance of existing models. A hierarchical neural network based on QRWs (QWNN) is developed by \cite{dernbach2018quantum} with a series of coins. The different setting of coins allows the quantum walks to behave differently in terms of extracting various properties of graphs. A classical diffusion process is employed to intersperse the information along with the composite form of position distribution generated by different quantum walks. Another quantum information propagation approach presented by \cite{bai2021learning} captures the multi-scale node features by employing the mixing matrix of CT-QRWs instead of directly using the adjacency matrix. QRWs are employed to select the more relevant neighbors of the center node to achieve a more efficient information diffusion process in the quantum subgraph convolutional neural network (QSCNN) \cite{zhang2019quantum}. \cite{chen2021hybrid} develops a composite model (QGCNN) to perform convolutional and pooling operations on graphs in which the parameterized quantum circuit tailed by the fully connected neural network is used to approximate the learning function. A hybrid GNN (HQGNN) is applied for particle track reconstruction \cite{tuysuz2021hybrid}, consisting of variational circuits to improve expressiveness. 
%It is reasonable to believe that the hybrid models have broad prospects, as the full computational power of quantum computing has not been fully tapped so far.
\vspace{-3pt}
\section{Challenges and Outlook}
\vspace{-3pt}
Though quantum graph computing and quantum graph learning have proven their potential in various learning tasks, challenges still exist due to the restricted scale of current quantum computational devices, instability of quantum states and complexity of yielding numerical results by measurement. We outlook three directions for the future researches.

%The structure information plays an important role in solving graph-related tasks. 
\textbf{Encoding Reliability} It is necessary to develop an effective approach to encode the graph into quantum representation while preserving the structural information and node’s and edge’s features. Besides, the success of (deep) neural networks relies on the nonlinear activation functions to enhance the expression ability, whereas the evolution of the quantum physics is linearly unitary transformation \cite{schuld2018supervised}. Therefore, it is also important to consider more nonlinear effects while embedding classical data.

\textbf{Data-driven and Adaptivity} The goal of developing classical learning methods is to be independent of experts and easy to be migrated to other tasks, and the same is true for quantum machine learning. However, most existing quantum computing methods are theory-oriented and knowledge-driven. A recent study \cite{huang2021power} demonstrates the potential of enhancing quantum algorithms using the power of data. The further study of data-driven quantum graph algorithms will be possible to provide a more valuable reference for solving graph theoretical and graph learning problems.

\textbf{Computational Efficiency}
The cost landscapes of training quantum algorithms are generally non-convex, making it difficult to establish general guarantees about the computational expense of the optimizations \cite{cerezo2020variational}. The process of embedding structural information into quantum expression will inevitably introduce additional ancilla qubits \cite{zheng2021quantum} and entanglement between qubits \cite{verdon2019quantum}, which further makes quantum circuits more difficult to model. These problems may be avoided by ingenious circuit design and efficient gradient calculation. 
%In a nutshell, efficiency should be considered when developing quantum algorithms.

%% The file named.bst is a bibliography style file for BibTeX 0.99c
\small
\bibliographystyle{named}
\bibliography{shorter}

\end{document}